\pdfoutput=1
\documentclass[runningheads]{llncs}
\usepackage[T1]{fontenc}
\usepackage{graphicx}
\usepackage{color}
\usepackage{cite}
\usepackage{subfigure}
\usepackage{booktabs}

\usepackage[utf8]{inputenc}
\usepackage[marginal]{footmisc}

\begin{document}
\title{U-Net Based Healthy 3D Brain Tissue Inpainting\thanks{Corresponding author: Ying Weng.}}

%
%
\author{Juexin Zhang\orcidID{0000-0001-9086-7342} \and
Ying Weng\orcidID{0000-0003-4338-713X} \and
Ke Chen\orcidID{0000-0002-2046-0034}}
\authorrunning{J. Zhang et al.}
%
\institute{University of Nottingham Ningbo China, Ningbo 315100, China
\email{ying.weng@nottingham.edu.cn}\\}
\maketitle              
\begin{abstract}

This paper introduces a novel approach to synthesize healthy 3D brain tissue from masked input images, specifically focusing on the task of `ASNR-MICCAI BraTS Local Synthesis of Tissue via Inpainting'. Our proposed method employs a U-Net-based architecture, which is designed to effectively reconstruct the missing or corrupted regions of brain MRI scans. To enhance our model's generalization capabilities and robustness, we implement a comprehensive data augmentation strategy that involves randomly masking healthy images during training. Our model is trained on the BraTS-Local-Inpainting dataset and demonstrates the exceptional performance in recovering healthy brain tissue. The evaluation metrics employed, including Structural Similarity Index (SSIM), Peak Signal-to-Noise Ratio (PSNR), and Mean Squared Error (MSE), consistently yields impressive results. On the BraTS-Local-Inpainting validation set, our model achieved an SSIM score of 0.841, a PSNR score of 23.257, and an MSE score of 0.007. Notably, these evaluation metrics exhibit relatively low standard deviations, i.e., 0.103 for SSIM score, 4.213 for PSNR score and 0.007 for MSE score, which indicates that our model's reliability and consistency across various input scenarios. Our method also secured first place in the challenge.

\end{abstract}

\section{Introduction}

Brain tumors, a devastating type of cancer, pose a significant challenge to healthcare providers worldwide. These tumors, characterized by the abnormal growth of cells within the brain, often lead to severe neurological symptoms and can be fatal if left untreated. The complexity of the brain's structure and the delicate nature of brain surgery make the treatment of brain tumors particularly demanding.

The diagnosis and management of brain tumors heavily rely on medical imaging techniques, with magnetic resonance imaging (MRI) being the most commonly used modality. MRI provides detailed images of the brain's anatomy and can help identify the location, size, and grade of tumors. However, interpreting MRI scans can be challenging, especially for non-experts, and can lead to diagnostic errors.

In recent years, artificial intelligence (AI) has emerged as a promising tool for improving the diagnosis and treatment of brain tumors. AI-based algorithms can analyze MRI scans more efficiently and accurately than humans, potentially leading to earlier detection and more effective treatment. For example, AI can be used to identify subtle abnormalities in MRI scans that may be missed by human radiologists, and to predict the growth rate and aggressiveness of tumors.

Patient privacy regulations pose a significant barrier to the advancement of AI-based brain tumor analysis, as they limit the availability of large, diverse datasets necessary for developing robust and accurate models. Brain tumor datasets are often small and heterogeneous, making it difficult to train robust and generalizable models. The typical sequence of image acquisition often starts with scans that already show signs of pathology. This can introduce bias into the image pre-processing stage and may not provide a complete representation of the lesion \cite{kofler2023brain}. Directly using brain pathological images can influence pre-processing steps like brain registration, skull removal, and tissue segmentation, which can impact the accuracy of AI-based predictions and clinical decision-making.

One promising approach to address the data scarcity problem is to synthesize healthy brain tissue from pathological MRI scans. This technique, known as inpainting, can be used to create synthetic training data that is more representative of healthy brains, thereby improving the performance of AI models. In this paper, we introduce a U-Net-like model that is specifically designed for inpainting brain tumor regions. Our model takes as input a pathological MRI scan and synthesizes healthy brain tissue in the affected area. The synthesized tissue is then used to create a synthetic healthy scan that can be used for training AI models.

The remainder of this paper is structured as follows: The BraTS dataset and the methodologies related to the U-Net like model are described in Section \ref{methods}. Section \ref{results} presents the experimental methods of the proposed model, and the paper is concluded in Section \ref{conclusion}.

\section{Methods}\label{methods}
\subsection{Dataset}

The BraTS-Local-Inpainting dataset (BraTS 2023) \cite{kofler2023brain} was employed for training, validation, and testing our model. This dataset exclusively comprises T1 MRI scans from the BraTS-GLI 2023 dataset \cite{baid2021rsna}. It includes 1251 scans, each accompanied by expert neuroradiologist-approved tumor region annotations. A sophisticated mask generation algorithm, as described in \cite{kofler2023brain}, accurately delineated healthy brain tissue regions that were spatially separated from the tumor. To prevent overfitting and improve model generalization, healthy masks were augmented through random mirroring and rotation. The resulting MRI volumes and associated masks have dimensions of 240 × 240 × 155. The training set contains four distinct data types:

\begin{itemize}
\item t1n: The ground truth image, which includes all tissue.
\item t1n-voided: Images that have had the healthy and unhealthy tissue removed.
\item healthy mask: A mask that identifies part of the healthy tissue in the t1n image.
\item unhealthy mask: A mask that identifies the tumor region in the t1n image.
\item mask: A mask that combines the healthy and unhealthy masks.
\end{itemize}

\subsection{Pre-processing}

The BraTS 2022 dataset underwent standard pre-processing steps, including aligning images to a common anatomical template, resampling to a uniform resolution of $1mm^3$ resolution, and removing the skull. Our approach involves additional cropping of MRI scans and masks to a size of 128 x 128 x 96. Importantly, we combine our model's predictions on cropped patches with the original t1-weighted MRI image using a stitching technique to generate the final output. Initially, we normalize the images to a range of $[0,1]$ by dividing each image by its maximum value, followed by normalization to a range of $[-1,1]$.

\begin{figure}
    \centering
    \includegraphics[width = \textwidth]{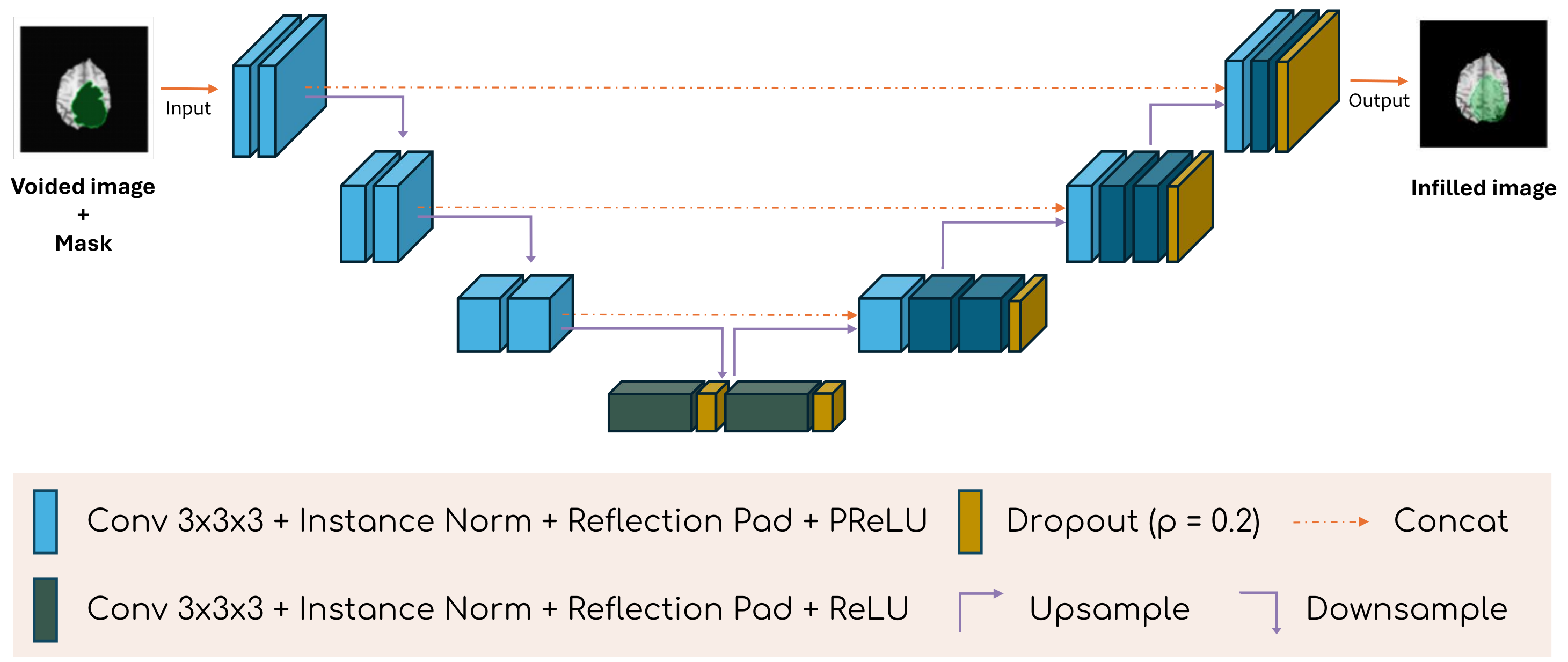}
    \caption{Architecture of our U-Net model. It is important to note that our model processes cropped images as both input and output, which differs slightly from the visualization in the figure. Typically, these cropped images encompass the entire brain tissue.}
    \label{unet}
\end{figure}

\subsection{Data Augmentation}

Deep learning models with extensive parameters often suffer from overfitting, hindering their ability to generalize to unseen data. To mitigate this issue, we generate five unique healthy masks for each MRI scan using the healthy inpainting mask generation algorithm described in \cite{kofler2023brain}. These masks are further augmented with random mirroring and rotation. Random mirroring is applied independently to each dimension with a 50\% probability, and a random angle between 0° and 360° is used for both the X-Y and Y-Z planes. While some overlap between different healthy masks for each MRI scan is expected, the augmented data helps train a more robust and generalized model due to the variability in mask shape, location, and size.

\subsection{Network Architecture}

We propose a U-Net \cite{ronneberger2015u} based model aimed at synthesizing healthy tissues. The architecture of our U-Net is illustrated in Figure \ref{unet}. Our model comprises three types of blocks: downsampling blocks, bridge blocks, and upsampling blocks. Each block includes two 3D convolutional layers with a kernel size of 3. ReLU activation functions are used in the bridge block, while PReLU is employed in the downsampling and upsampling blocks. Instance normalization is applied to normalize the data for each MRI scan. To prevent overfitting, both the bridge block and upsampling block incorporate dropout. Our model features three downsampling/upsampling blocks and one bridge block. The initial downsampling block has 32 channels, with the number of channels doubling in each subsequent downsampling block. Conversely, the number of channels in each upsampling block is half that of the corresponding downsampling block. Skip connections are utilized to transmit features from different levels of the downsampling blocks to the upsampling blocks. Our model accepts t1n-voided images and masks as input and generates infilled images as output. Two loss functions are applied to train our model: mean absolute error (MAE) and structural similarity index measure (SSIM) \cite{wang2004image}. The MAE is calculated exclusively between the healthy regions of the ground truth image ($GT$) and the infilled image ($I$), while SSIM is computed for the entire images. We observed that the SSIM loss significantly underperforms MAE loss in preserving mask regions. Additionally, the structural similarity focus of SSIM may overlook mask edges, resulting in generated content that mismatches the mask boundaries. The loss functions are defined as follows:"

\begin{equation}
    MAE(x, y) = \frac{1}{m}\sum_{i=1}^{m}|y_{i}-f(x_{i})|
\end{equation}
\begin{equation}
    SSIM(x, y) = \frac{(2\mu_{x}\mu_{y}+c_{1})(2\sigma_{xy}+c_{2})}{(\mu_{x}^{2}+\mu_{y}^{2}+c_{1})(\sigma_{x}^{2}+\sigma_{y}^{2}+c_{2})}
\end{equation}
\begin{equation}
    Loss(I, GT) = \lambda_{1} \times MAE(I, GT) + \lambda_{2} \times SSIM(I, GT)
\end{equation}

\section{Experiment Results}\label{results}
\subsection{Evaluation Metrics}
To measure the performance of our model, we evaluate the generated healthy regions using three metrics: structural similarity index measure (SSIM), peak-signal-to-noise-ratio (PSNR), and mean-square-error (MSE). It is worth noting that we only the healthy regions defined by healthy masks against the ground truth data.

\subsection{Experiment Settings}
In our experiments, we use 5-fold cross-validation to select hyperparameters during training.  We set the maximum number of epochs to 500 and saved checkpoints for the five models with the lowest validation loss. We set the dropout rate equal to 0.2 in the bridge block and upsample block. In validation phase, we applied Adam optimizer with initial learning rate $1e^{-4}$ and betas $(0.9, 0.999)$ and we set $\lambda_{1}$ and $\lambda_{2}$ to 1 in our loss function. We normalize the healthy mask region to a range of [0, 1] based on the maximum value of the healthy and unhealthy regions of the ground truth image in validation phase. The model is implemented with PyTorch and all experiments are conducted on a single Nvidia RTX 3090 with 24GB VRAM.

\subsection{Validation Phase}

\begin{figure}
    \centering
    \subfigure[BraTS-GLI-01717-000 (best)]{
    \includegraphics[width=\textwidth]{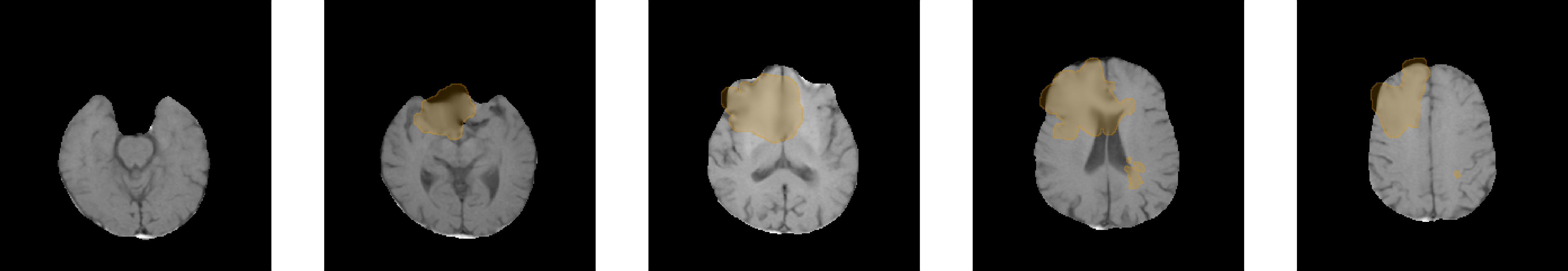}\label{sub1}}
    \subfigure[BraTS-GLI-01739-000 (median)]{
    \includegraphics[width=\textwidth]{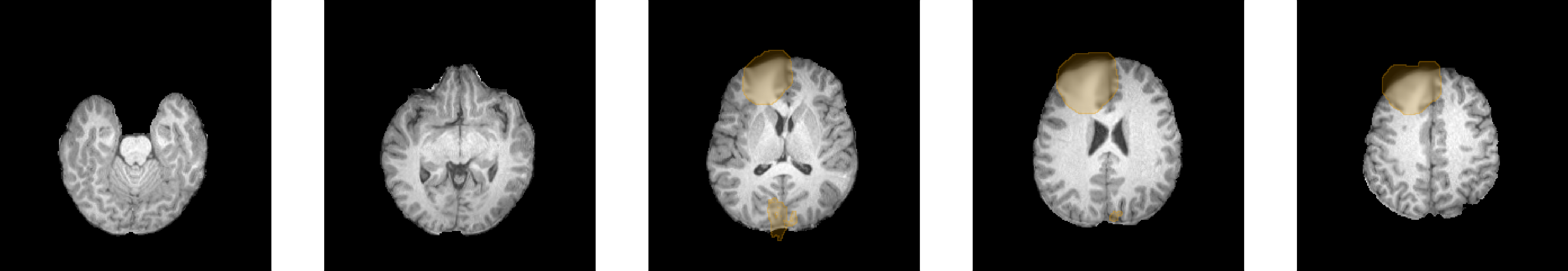}\label{sub2}}
    \subfigure[BraTS-GLI-00671-000 (worst)]{
    \includegraphics[width=\textwidth]{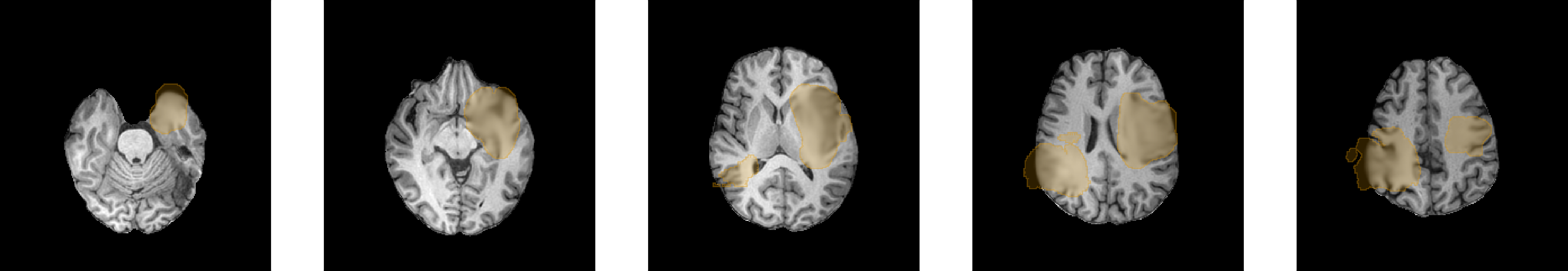}\label{sub3}}

    \caption{Visualization of the infilled validation MRI scans. In the images, the orange portions represent the masked regions containing both healthy and unhealthy tissues, since they were not explicitly labeled in the validation dataset. During the validation phase, our model's performance has been exclusively assessed using healthy tissue segments. Fig \ref{sub1} has achieved an SSIM score of 0.999154925, a PSNR score of 37.65698624 and an MSE score of 0.000171515. Fig \ref{sub2} has achieved an SSIM score of 0.929092348, a PSNR score of 22.3681049 and an MSE score of 0.005796814. Fig \ref{sub3} has achieved an SSIM score of 0.647443771, a PSNR score of 16.45079041 and an MSE score of 0.022642322. In Fig \ref{sub2} and \ref{sub3}, we selected the scans with the median and lowest values of MSE and PSNR scores, which are not corresponding to the median and lowest values of SSIM scores.}
    \label{fig1}
\end{figure}

We present the overall results of our model on the BraTS-Local-Inpainting validation dataset in Table \ref{tab1}. The evaluation metric values were computed by the online evaluation platform of Sage Bionetworks Synapse (Synapse). We also visualize the best, median, and worst infilled results of our model on the validation set in Figure \ref{fig1}. However, the ground truth data is not visible in the validation phase or test phase, so we cannot visualize the ground truth data to compare with our infilled images. Additionally, the orange areas are mask areas that include both healthy and unhealthy tissues, but our aim is to only infill the healthy tissues, so only the healthy areas are evaluated. Visual analysis reveals our model's ability to effectively capture low-level textures and synthesize brain tissues. The structure of the infilled areas closely resembles the surrounding regions. However, the infilled areas exhibit blurriness, particularly in low-intensity regions (the fourth image in Figure \ref{sub2} and the first image in Figure \ref{sub3}). This blurriness may be attributed to the MAE loss, which averages the error across the entire image. This can lead to the model smoothing the image to minimize the error, even at the cost of detail.
 
Table \ref{tab1} presents the overall performance of our model on the BraTS-Local-Inpainting validation dataset. Evaluation metrics were calculated using the online evaluation platform of Sage Bionetworks Synapse (Synapse). The best, median, and worst infilled results of our model on the validation set are visualized in Figure \ref{fig1}. Due to the nature of the validation and test phases, ground truth data is not visible, preventing direct comparison with our infilled images. The orange areas in the masks represent both healthy and unhealthy tissues, but our focus is on infilling healthy tissues only, hence the evaluation is restricted to healthy regions.

\begin{table}[ht]
\centering
\caption{Overall validation data results}
\label{tab1}
\begin{tabular}{lcccccc}
\toprule
 & MSE & PSNR & SSIM \\
\midrule
Mean & 0.006503617 & 23.381424612106255 & 0.84116631773508840\\
Standard deviation & 0.004660640 & 4.2644961127908990 & 0.10317845134515838\\
25 quantile & 0.002789088 & 20.387409210205078 & 0.75842395424842830\\
Median & 0.005796814 & 22.368104934692383 & 0.84412175416946410\\
75 quantile & 0.009146666 & 25.545470237731934 & 0.92018622159957890\\

\bottomrule
\end{tabular}
\end{table}

\section{Conclusion\label{conclusion}}

This paper presents our contributions to the task of `ASNR-MICCAI BraTS Local Synthesis of Tissue via Inpainting'. A U-Net like model has been proposed to synthesize the healthy 3D brain tissue. Multiple data augmentations have been applied, including random healthy mask generation, random  mirroring and rotation. Our derived model is trained on the BraTS-Local-Inpainting training set and its performance on the BraTS-Local-Inpainting validation set computed by the online evaluation platform Synapse has been shown in Table \ref{tab1}. Remarkably, our method has achieved an SSIM score of 0.841, a PSNR score of 23.257 and an MSE score of 0.007. Meanwhile, our model also has a relatively low standard deviation for the three evaluation metrics, i.e., 0.103 for SSIM score, 4.213 for PSNR score and 0.007 for MSE score.

\section*{Acknowledgements}
This work was supported by Ningbo Major Science \& Technology Project under Grant 2022Z126. (Corresponding: Ying Weng.)

%
%
%
\bibliographystyle{splncs04}
\bibliography{ref.bib}

\end{document}